\documentclass[preprint,12pt]{elsarticle}
\journal{Journal of Modern Optics}
\usepackage{amsfonts,physics}
\usepackage[utf8]{inputenc}
\usepackage{hyperref}
\usepackage{xcolor}
\hypersetup{colorlinks=true,urlcolor=red,citecolor=black,linkcolor=black}
\allowdisplaybreaks
\usepackage{amsmath}
\allowdisplaybreaks
\usepackage{amsfonts}
\usepackage{amssymb}
\usepackage{graphicx}
\usepackage{float}
\newcommand{\bluecite}[1]{\textcolor{black}{\cite{#1}}}
\begin{document}
	\title{Dynamics of atom-field
			interaction inside a nonlinear Kerr-like medium filled optical cavity}
	\author{Naveen Kumar}
	\ead{naveen74418@gmail.com}
	\author{Arpita Chatterjee*}
	\ead{arpita.sps@gmail.com}
	\cortext[cor1]{Corresponding author}
	\date{\today}
	\address{Department of Mathematics, J. C. Bose University of Science and Technology,\\ YMCA, Faridabad 121006, India}
	\begin{abstract}
		
		In this paper, we investigate the dynamics of two  two-level atoms interacting with a two-mode field inside an optical cavity, in presence of a nonlinear Kerr-like medium as well as the Stark shift. We derive the exact analytical solution of the time-dependent Schrödinger equation that provides a comprehensive framework for analyzing the system's quantum properties. To characterize the nonclassical features of the radiation field, we examine photon number distribution, second-order correlation function $g^2(0)$, squeezing properties, and Mandel's $Q_M$ parameter. These properties reveal significant insights into the quantum statistical behaviour of the field and its deviation from classicality under different interaction regimes. In addition, we quantify the atom-atom entanglement using linear entropy which captures the mixedness of the atomic subsystem and elucidates the interplay between atom-atom interactions. The results highlight the crucial role of nonlinear interactions and the Stark shift in shaping the quantum correlations and nonclassical phenomena of the system.
		\end{abstract}
	\maketitle

	\section{Introduction}
	The Jaynes-Cummings model (JCM) \cite{Jaynes1963} is the fascinating model in the field of quantum optics and cavity quantum electrodynamics (QED) that explains the interaction between a two-level atom and a single-mode of a quantized radiation field when the rotating-wave approximation (RWA) is considered. This theoretical model can be solved analytically and is essential for understanding how a two-level atom interacts with a single-mode quantized cavity electromagnetic field and the dynamics that emerges from their coupling. Over the years, researchers have investigated numerous extensions of the original JCM, broadening its scope beyond the two-level atom and single-mode cavity interaction. These extensions involve multi-level atoms and multi-mode cavities, leading to a rich variety of phenomena and applications in quantum optics and cavity QED \cite{shore1993jaynes}.
	
	Significant attention has been paid to obtain the analytical solution of the system that involves multiple atoms interacting with the cavity field. An interesting example is investigating the dynamic behaviour of two two-level atoms with a single-mode field \cite{jex1992emission,
		bougouffa2010entanglement,
		ashraf1999effects,hekmatara2014sub}. This scenario provides a major example that explores the complex dynamics arising from the collective interaction among atoms and the cavity electromagnetic field. A number of studies have been reported on atom-atom and atoms-field entanglement \cite{zidan2002influence, el2003entropy}. In recent years, multiple extensions of the JCM such as multi-photon transitions, multi-mode fields, intensity-dependent coupling, parametric down-conversion, Kerr-medium phenomena, cross-Kerr medium interactions, and Stark shifts  have been investigated \cite{baghshahi2015entanglement,ateto2009control,
hu2008effect,baghshahi2014entanglement}. The generation and characterization of nonclassical and entangled states represent a frontier of research in quantum physics, offering profound insights into fundamental principles and practical applications \cite{aspect1982experimental}. Specifically, the interaction between two two-level atoms and a two-mode field within an optical cavity presents a unique and intriguing scenario for studying such quantum states. The Jaynes-Cummings model serves as a foundational framework to understand the dynamics of such systems. The coupling between the atoms and the field leads to the emergence of nonclassical states, characterized by the properties that defy classical descriptions. Moreover, the inherent correlations between the atoms and the field give rise to entangled states, highlighting the interconnectedness between quantum entities \cite{nielsen2010quantum}. 

In recent years, notable advancement is seen in the theoretical and experimental investigation of nonclassical and entangled resources presented in Kerr-like medium filled optical cavities \cite{haroche2006exploring}. These studies have unveiled potential applications of such states in quantum information processing, quantum communication, and quantum metrology. By harnessing the unique properties of nonclassical and entangled states in Kerr-like medium filled optical cavities, we aim to unlock new frontiers in quantum information processing and quantum technology. Also when multiple atoms interact with a cavity field, the system exhibits heightened complexity rising to multipartite entanglement and unique quantum correlations. Understanding the dynamics of such systems is paramount for advancing scalable quantum technologies and implementing quantum algorithms that surpass their classical counterparts \cite{imamoglu1997strongly,molmer1999multiparticle}. 

	
	In this paper, we aim to investigate the dynamics of two two-level atoms interacting with a two-mode field in presence of Kerr-like medium and detuning parameter. We consider that the atom-field coupling is
$f$-deformed (intensity-dependent regime). We use the adiabatic elimination method to simplify the Hamiltonian of the entire atom-field system. 
Furthermore, we derive the explicit form of the state vector for the entire system by solving the time-dependent Schrödinger equation. Our analysis reveals the entanglement properties inherent in the system, showcasing the intricate atom-field and atom-atom quantum correlations \cite{poyatos1997complete}. 
	
	\section{State of interest}
	In quantum mechanics, the wave function of a physical system contains extensive information about the system's properties and behaviour. Thus, the primary step for analyzing such a system is to obtain the state vector explicitly that requires a thorough understanding of all the interactions that influence the system and lead to the construction of an appropriate Hamiltonian. By identifying and characterizing the system's interactions, one can formulate a Hamiltonian that accurately describes its dynamics. This Hamiltonian serves as the mathematical framework to determine the time evolution of the system's state vector using the Schrödinger equation. Obtaining the explicit form of the state vector enables the exploration and interpretation of the system's quantum phenomena that provides valuable insights about its observable characteristics and possible outcomes. 
	
	Here we consider a system where two two-level atoms (atom $A$ and atom $B$ with ground states $\ket{g_A}$, $\ket{g_B}$ and excited states $\ket{e_A}$, $\ket{e_B}$, respectively) interact with a two-mode quantized field with frequencies $\Omega_1$ and $\Omega_2$ in an optical cavity. The cavity is filled by a centrosymmetric medium featuring Kerr nonlinearity under the impact of detuning parameter and Stark shift. We provide a schematic representation for the atom-field interaction. In a cavity QED set up involving a Kerr medium, the nonlinear crystal constituting the Kerr medium is positioned within the optical cavity. This cavity supports two quantized modes of electromagnetic radiation that interact with the medium. The refractive index of the Kerr medium depending on the field intensity leads to a nonlinear phase shift that modifies the dynamics of the quantized field modes. The two two-level atoms, initially prepared in superposition states, are confined in fixed positions within the cavity. These atoms are trapped using external mechanisms such as optical tweezers or magnetic fields, ensuring precise localization and interaction with the quantized field. The quantized field interacts with both the atoms and the Kerr medium simultaneously. The coupling between the field and atoms is described by an interaction Hamiltonian that incorporates intensity-dependent terms. These terms account for the influence of the Kerr medium, Stark shifts, and possible detuning effects. While the atoms do not physically move through the Kerr medium, the cavity field altered by the nonlinear properties of the Kerr medium, mediates the interaction between the atoms and the quantized modes. This configuration enables the study of quantum properties such as entanglement, coherence, and population inversion \cite{korashy2020dynamics,abdel1992degenerate,miranowicz2004dissipation}.
	
	\begin{figure}[htb]
		\centering
		\includegraphics[width=0.7\textwidth]{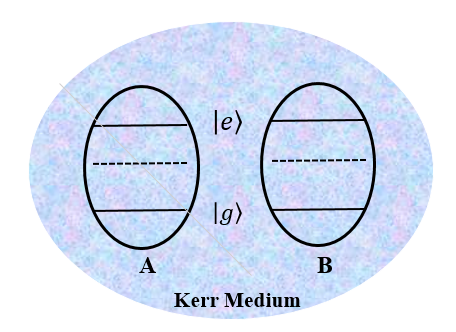}
		\caption{A schematic representation of the atom-field interaction within an optical cavity incorporating a medium with Kerr nonlinearity.}
		\label{int}
	\end{figure}
	
In the rotating-wave approximation, the Hamiltonian representing all the interactions of the considered physical system is ($\hbar=1$)	\begin{align}
		\nonumber
		H&=\frac{1}{2}\sum_{i=A,B}\omega_i\sigma_z^i+\sum_{j=1}^{2}\Omega_ja_j^\dag a_j+\sum_{i=A,B}g_i(t)(R_1R_2\sigma_+^i+R_1^\dag R_2^\dag\sigma_-^i)\\
		&+\sum_{i=A,B}(\beta_1^ia_1^\dag a_1\sigma_-^i \sigma_+^i+\beta_2^ia_2^\dag a_2\sigma_+^i \sigma_-^i)+\sum_{j=1}^{2}\chi_j a_j^{\dag^2}a_j^2+\chi_{12} a_1^\dag a_1 a_2^\dag a_2
		\label{h1}
	\end{align}
	Here $\frac{1}{2}\sum_{i=A,B}\omega_i\sigma_z^i$ represents the energy associated with the two atoms labelled as $A$ and $B$. $\omega_i$ $(i=A, B)$ denotes the atomic transition frequency associated with the energy difference between the two levels of the atoms that means $\omega_i$ essentially accounts the energy difference between the ground and excited states of each atom in the system. The second part of the Hamiltonian, $\sum_{j=1}^{2}\Omega_ja_j^\dag a_j$ signifies the energy associated with the quantized electromagnetic fields in two distinct modes ($j = 1, 2$). Each mode is characterized by its angular frequency $\Omega_j$. The operators $a_j^\dagger$ and $a_j$ are the field creation and annihilation operators, respectively, for photons in $j$-th mode. This term accounts for the energy of the photons present in each mode where the number of photons can vary due to the action of the creation and annihilation operators. $\sum_{i=A,B}g_i(t)(R_1R_2\sigma_+^i+R_1^\dag R_2^\dag\sigma_-^i)$ describes the interaction between the atoms and the two modes of the electromagnetic field. The function $g_i(t)$ represents the time-dependent coupling strength between $i$-th atom and the electromagnetic field modes. These terms capture the energy exchange between the atoms and the electromagnetic field modes through absorption and emission processes. 
	The operators $\sigma_z^i$, $\sigma_\pm^i$ are the atomic pseudospin operators corresponding to the $i$-th two-level atom. The parameters $\beta_1^i$ and $\beta_2^i$ characterize the effective Stark shift coefficients for the atoms. Furthermore, $\chi_j$ and $\chi_{12}$ show the cubic susceptibility of the medium, where $\chi_j$ signifies the Kerr self-action for mode $j$, and $\chi_{12}$ is related to the cross-Kerr process. In addition, the nonlinear ($f$-deformed) annihilation operator is introduced by $R_j=a_jf({N_j})$ and the corresponding creation operator is denoted by $R_j^\dag=f({N_j})a_j^\dag$ with $N_j=a_j^\dagger a_j$ and $f({N_j})$ represents the Hermitian operator-valued functions responsible for the intensity-dependent atom-field coupling. These operators adhere to certain commutation relations as following:\\
	$$[R_j, N_j]=R_j  ~ ~\text{and}~ ~ [R_j^\dag, N_j]=-R_j^\dag$$
	For the subsequent analysis, it is advantageous to rewrite the Hamiltonian (\ref{h1}) in the interaction picture. The Hamiltonian \eqref{h1} can be written in following manner:
	\begin{equation}
		H=H_0+H_I
	\end{equation}
	where 
	\begin{align*}
		\nonumber
		H_0&=\sum_{i=A,B}\omega_1\sigma_z^i+\frac{1}{2}\sum_{j=1}^{2}\Omega_ja_j^\dag a_j\\
		H_I&=
		\sum_{i=A,B}g_i(t)(R_1R_2\sigma_+^i+R_1^\dag R_2^\dag \sigma_-^i)+\sum_{i=A,B}(\beta_1^ia_1^\dag a_1 \sigma_{gg}^i+\beta_2^ia_2^\dag a_2\sigma_{ee}^i)\\
		&+\sum_{j=1}^{2}\chi_j a_j^{\dag^2}a_j^2+ \chi_{12} a_1^\dag a_1 a_2^\dag a_2
	\end{align*}
	Now applying the Baker–Campbell–Hausdorff (BCH) formula
	$$e^{iH_0t}H_I e^{-iH_0t}=H_I+it[H_0,H_I]+\frac{{(it)}^2}{2}[H_0,[H_0,H_I]]+\ldots,$$
	the effective Hamiltonian can be written as
	\begin{align}
		\nonumber
		H_{\text{eff}}&=e^{iH_0t}H_Ie^{-iH_0t}\\\nonumber
		&=\frac{1}{2}\sum_{i=A,B}\Delta_i\sigma_z^i+\sum_{i=A,B}g_i(t)\left(R_1R_2\sigma_+^ie^{-i \Delta_i t}+R_1^\dag R_2^\dag \sigma_-^ie^{i\Delta_i t}\right)\\
		&+\sum_{i=A,B}\left(\beta_1^ia_1^\dag a_1 \sigma_{gg}^i+\beta_2^ia_2^\dag a_2\sigma_{ee}^i\right)+\sum_{j=1}^{2}\chi_j a_j^{\dag^2}a_j^2+ \chi_{12} a_1^\dag a_1 a_2^\dag a_2
		\label{heff}
	\end{align}
	where $\Delta_i=\omega_i-\Omega_1-\Omega_2$ represent the detuning parameters. For simplicity and enhanced realism, let us assume that the atoms are identical. Thus we have $\Delta_A=\Delta_B=\Delta$, $\beta_1^A=\beta_1^B=\beta_1$, $\beta_2^A=\beta_2^B=\beta_2$, and $g_A(t)=g_B(t)=g(t)$.
	The inclusion of the time-dependent coupling parameter renders the entire Hamiltonian time-dependent. Different forms can be chosen for $g(t)$, and in this context we opt for $g(t)=\lambda\cos(\epsilon t + \phi)$, where $\lambda$ and $\epsilon$ are arbitrary constants and $\phi$ represents the relative phase.
	
	We assume that the wave function describing the atom-field system at any time $t>0$ can be formulated as:
	\begin{align}
		\nonumber
		&\ket{\psi(t)}\\\nonumber&=\sum_{n_1,n_2=0}^{\infty}q_{n_1}q_{n_2}\bigg[A_1(n_1,n_2,t)\ket{e_1,e_2,n_1,n_2}+A_2(n_1+1,n_2+1,t)\ket{e_1,g_2,n_1+1,n_2+1}\\
		&+A_3(n_1+1,n_2+1,t)\ket{g_1,e_2,n_1+1,n_2+1}+A_4(n_1+2,n_2+2,t)\ket{g_1,g_2,n_1+2,n_2+2}\bigg]
		\label{psi}
	\end{align}
	In \eqref{psi}, $q_{n_1}$ and $q_{n_2}$ are the probability amplitudes of the initial states of the radiation fields of the cavity. The atomic probability amplitudes to be determined are denoted by $A_1$, $A_2$, $A_3$, and $A_4$.
	
	Solving the time-dependent Schrödinger equation $i\frac{\partial}{\partial t}\ket{\psi(t)}=H_{\text{eff}}\ket{\psi(t)}$ \cite{scully1997quantum}, one can find out the coupled differential equations in terms of the probability amplitudes of the state vector $\ket{\psi(t)}$ as 
	\begin{align}
		\nonumber
		i\dot{A_1}(n_1,n_2,t)&=T_1 A_1(n_1, n_2, t)+2g(t)e^{i\Delta t}V_1A_2(n_1+1, n_2+1, t)\\\nonumber
		i	\dot{A_2}(n_1+1,n_2+1,t)&=T_2A_2(n_1+1, n_2+1, t)+g(t)e^{-i\Delta t}V_1A_1(n_1,n_2,t)\\
		&+g(t)e^{i\Delta t}V_2A_4(n_1+2,n_2+2,t)\\\nonumber
		i	\dot{A_4}(n_1+2,n_2+2,t)&=T_4A_4(n_1+2,n_2+2,t)+2g(t)e^{-i\Delta t}V_2A_2(n_1+1, n_2+1, t)
		\label{prob}	
	\end{align}
	where 
	\begin{align}
		\nonumber
		& A_2(n_1+1,n_2+1,t)=A_3(n_1+1,n_2+1,t)\\\nonumber
		&V_1=f(n_1+1)f(n_2+1)\sqrt{(n_1+1)(n_2+1)}\\
		&V_2=f(n_1+2)f(n_2+2)\sqrt{(n_1+2)(n_2+2)}\\\nonumber
		&T_1=\Delta+2\beta_2 n_2+\chi_1n_1(n_1-1)+\chi_2n_2(n_2-1)+\chi_{12}n_1n_2\\\nonumber
		&T_2=\beta_1 (n_1+1)+\beta_2 (n_2+1)+\chi_1n_1(n_1+1)+\chi_2n_2(n_2+1)+\chi_{12}(n_1+1)(n_2+1)\\\nonumber
		&T_4=-\Delta+2\beta_1 (n_1+2)+\chi_1(n_1+1)(n_1+2)+\chi_2(n_1+1)(n_2+2)+\chi_{12} (n_1+2)(n_2+2)
	\end{align}
	To remove the fast frequency dependence of $A_1$, $A_2$, and $A_4$, we can transform them into slowly varying functions as $$A_1=C_1(t)e^{-i\,T_1t},A_2=A_3=C_2(t)e^{-i\,T_2t}~\text{and}~A_4=C_4(t)e^{-i\,T_4t},$$
	where $C_1(t)$, $C_2(t)$, and $C_4(t)$ are slowly varying functions of time and $T_1$, $T_2$, $T_4$ are associated with the fast frequencies that we aim to remove from the dynamics. This transformation simplifies the differential equations which govern the time evolution of the atomic probability amplitudes by separating the rapidly oscillating terms. Now \eqref{prob} can be further simplified into
	\begin{align}\nonumber
		i	\dot{C_1}(t)&=\lambda V_1\bigg[e^{i[\{\epsilon+(T_1-T_2)\} t+\phi]}+e^{-i[\{\epsilon-(T_1-T_2)\}t+\phi]}\bigg]C_2(t)\\\nonumber
		i	\dot{C_2}(t)&=\frac{\lambda}{2}\bigg[V_1\big\{e^{i[\{\epsilon-(T_1-T_2)\}t+\phi]}+e^{-i[\{\epsilon+(T_1-T_2)\}t+\phi]}\big\}C_1(t)\\\nonumber
		&+V_2\big\{e^{i[\{\epsilon+(T_2-T_4)\}t+\phi]}+e^{-i[\{\epsilon-(T_2-T_4)\}t+\phi]}\big\}C_4(t)\bigg]\\
		i	\dot{C_4}(t)&=V_2\bigg[e^{i[\{\epsilon-(T_2-T_4)\}t+\phi]}+e^{-i[\{\epsilon+(T_2-T_4)\}t+\phi]}\bigg]C_2(t)
		\label{de}
	\end{align}
	In any of the above equations, we can find the presence of two types of exponential terms: one exhibiting rapid oscillations such as $e^{i\{\epsilon+(T_1-T_2)\}t}$, $e^{i\{\epsilon+(T_2-T_4)\}t}$ while the other represents slowly varying terms like $e^{i\{\epsilon-(T_1-T_2)\}t}$, $e^{i\{\epsilon-(T_2-T_4)\}t}$. In this scenario, if we choose to ignore the rapidly oscillating terms as compared to the slowly varying terms, then \eqref{de} becomes
	\begin{align}
		\nonumber
		i	\dot{C_1}(t)&=\lambda V_1e^{-i[\{\epsilon-(T_1-T_2)\}t+\phi]}C_2(t)\\\nonumber
		i	\dot{C_2}(t)&=	\frac{\lambda}{2}\bigg[V_1e^{i[\{\epsilon-(T_1-T_2)\}t+\phi]}C_1(t)
		+V_2e^{-i[\{\epsilon-(T_2-T_4)\}t+\phi]}C_4(t)\bigg]\\
		i	\dot{C_4}(t)&=\lambda V_2 e^{i[\{\epsilon-(T_2-T_4)\}t+\phi]}C_2(t)
		\label{de1}
	\end{align}
	By substituting $C_4(t)=e^{imt}$ in \eqref{de1}, we can derive the following third-order algebraic equation as
	\begin{align}
		m^3+K_1m^2+K_2m+K_3&=0
		\label{roots}
	\end{align}
	where $a=\epsilon-(T_1-T_2)$, $b=\epsilon-(T_2-T_4)$, $K_1=-(a+2b)$, $K_2=b(a+b)-\frac{\lambda^2(V_1^2+V_2^2)}{2}$ and $K_3=\frac{\lambda^2(a+b)V_2^2}{2}$. In general, \eqref{roots} has three different roots which may be found as
	$$m_j=-\frac{1}{3}K_1+\frac{2}{3}\sqrt{K_1^2-3K_2}\cos(\Phi+ \frac{2}{3}(j-1)\pi),\,\,j=1, 2, 3$$ 
	$$\Phi=\frac{1}{3}\cos^{-1}\left[\frac{9K_1K_2-2K_1^3-27K_3}{2(K_1^2-3K_2)^{3/2}}\right]$$
	Considering $C_4(t)$ as a linear combination of three different $m_j$'s, the coefficients can be calculated as
	\begin{align}
		\nonumber
		C_1(t)&=\frac{1}{\lambda^2 V_1V_2}\sum_{k=1}^{3}\bigg[a_k(m_k^2-bm_k-\lambda^2 V_2^2)\bigg]e^{i[(m_k-a-b)t-2\phi]},\\\nonumber
		C_2(t)&=-\frac{1}{\lambda V_2}\sum_{k=1}^{3}a_km_ke^{i[(m_k-b)t-\phi]}\\
		C_4(t)&=\sum_{k=1}^{3}a_ke^{im_kt}
	\end{align}
	where $a_k$ can be determined by applying the initial conditions for both the atoms and the field. To achieve this, let us assume that the atoms enter the cavity in the coherent superposition of the states $\ket{e_1,e_2}$, $\ket{e_1,g_2}$, 
	$\ket{g_1,e_2}$, $\ket{g_1,g_2}$ that means
	$$\ket{\psi(0)}_{\text{atom}}=\gamma_1\ket{e_1,e_2}+\gamma_2\ket{e_1,g_2}+\gamma_3\ket{g_1,e_2}+\gamma_4\ket{g_1,g_2}$$
	where $\gamma_1$, $\gamma_2$, $\gamma_3$, $\gamma_4$ are arbitrary constants satisfying the normalization relation $\sum_{k=1}^4 |\gamma_k|^2=1$. Also, the field is assumed to be initially in the coherent state $$\ket{\psi(0)}_{\text{field}}=\sum_{n_1,n_2=0}^{\infty}q_{n_1}q_{n_2}\ket{n_1,n_2},$$
	with $q_{n_1}=e^{-|\alpha_1|^2/2}\frac{\alpha_1^{n_1}}{\sqrt{n_1!}}$ and $q_{n_2}=e^{-|\alpha_2|^2/2}\frac{\alpha_2^{n_2}}{\sqrt{n_2!}}$. By applying these initial conditions, $a_k$ becomes 
	$$a_k=\frac{\frac{\lambda^2 V_1V_2}{2}e^{2i\phi}\gamma_1+\lambda V_2(-b+m_p+m_q)e^{i\phi}\gamma_2+\left(\frac{\lambda^2 V_2^2}{2}+m_pm_q\right)\gamma_4}{(m_k-m_p)(m_k-m_q)},\,k\ne p\ne q=1,2,3$$
	The probability amplitudes can finally be obtained as 
	\begin{align*}
		A_1(n_1, n_2, t)&=\frac{1}{\lambda^2 V_1V_2}\sum_{k=1}^{3}a_k\left(m_k^2-bm_k-\lambda^2 V_2^2 \right) e^{i[(m_k-2\epsilon-T_4)t-2\phi]}\\
		A_2(n_1+1, n_2+1, t)&=-\frac{1}{\lambda V_2}\sum_{k=1}^{3}a_km_ke^{i[(m_k-\epsilon-T_4)t-\phi]}\\
		A_4(n_1+2, n_2+2, t)&=\sum_{k=1}^{3}a_ke^{i(m_k-T_4)t}
	\end{align*}
	After the atom-field interaction, the cavity field state is attained by tracing out the atomic part as
	$$\text{Tr}_\mathrm{atom}(\ket{\psi}\bra{\psi})=\ket{\psi_{\mathrm {field}}}\bra{\psi_{\mathrm {field}}}$$
	which is the state of interest throughout the rest of the article. Now the reduced density matrix for the field is
	\begin{align*}\nonumber
		&\rho_F\\&=\ket{\psi_{\mathrm {field}}}\bra{\psi_{\mathrm {field}}}\\
		&=\sum_{n_1,n_2=0}^{\infty}|q_{n_1}|^2|q_{n_2}|^2\bigg[|A_1|^2\ket{n_1,n_2}\bra{n_1,n_2}+2|A_2|^2\ket{n_1+1,n_2+1}\bra{n_1+1,n_2+1}\\
		&+|A_4|^2\ket{n_1+2,n_2+2}\bra{n_1+2,n_2+2}\bigg]
	\end{align*}
	\section{Generalized expectation}
	In the following section, we require certain expectation values with respect to the state of the cavity field. Consequently, the generalized expectation value for any two arbitrary positive integers $p$ and $q$ is derived as
	\begin{align*}
		&\bra{\psi_\mathrm{field}}a_1^{{\dag}^p} a_1^qa_2^{{\dag}^p} a_2^q\ket{\psi_
			\mathrm{field}}\\
		&=\sum_{n_1,n_2=0}^{\infty}q_{n_1+p-q}q_{n_2+p-q}q_{n_1}q_{n_2}\\
		&\times\bigg[A_1(n_1,n_2,t) A_1^*(n_1+p-q,n_2+p-q,t)\,\frac{\sqrt{n_1!n_2!(n_2+p-q)!(n_1+p-q)!}}{(n_1-q)!(n_2-q)!}\\
		&+4A_2(n_1+1,n_2+1,t) A_2^*(n_1+p-q+1,n_2+p-q+1,t)\\
		&\times\frac{\sqrt{(n_1+1)!(n_2+1)!(n_2+p-q+1)!(n_1+p-q+1)!}}{(n_1-q+1)!(n_2-q+1)!}\\
		&+A_4(n_1+2,n_2+2,t) A_4^*(n_1+p-q+2,n_2+p-q+2,t)\\
		&\times\frac{\sqrt{(n_1+2)!(n_2+2)!(n_2+p-q+2)!(n_1+p-q+2)!}}{(n_1-q+2)!(n_2-q+2)!}\bigg]
	\end{align*}
	
	\section{Photon number distribution}
\begin{figure}[htb]
		\centering
		\includegraphics[width=0.4\textwidth]{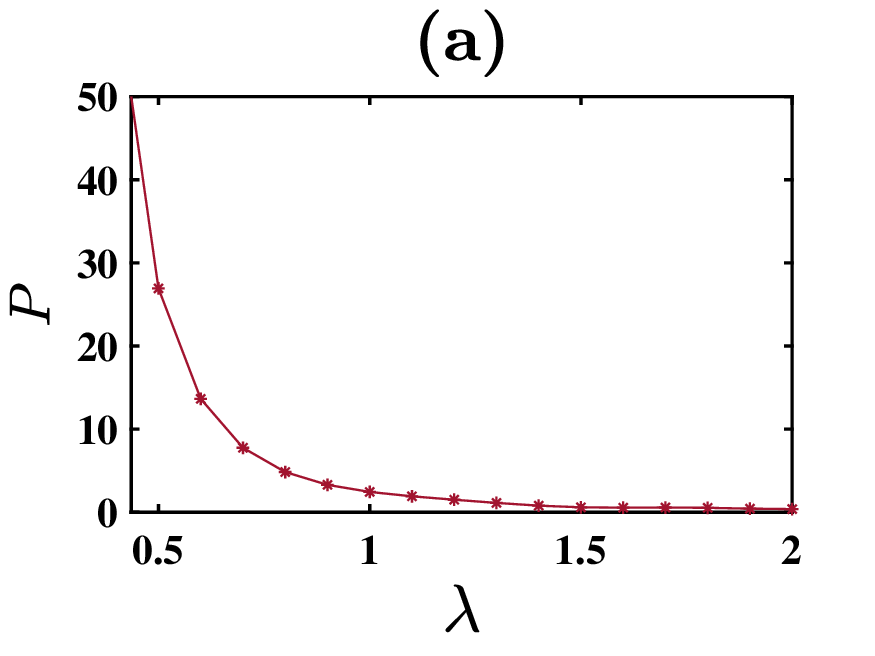}
		\includegraphics[width=0.4\textwidth]{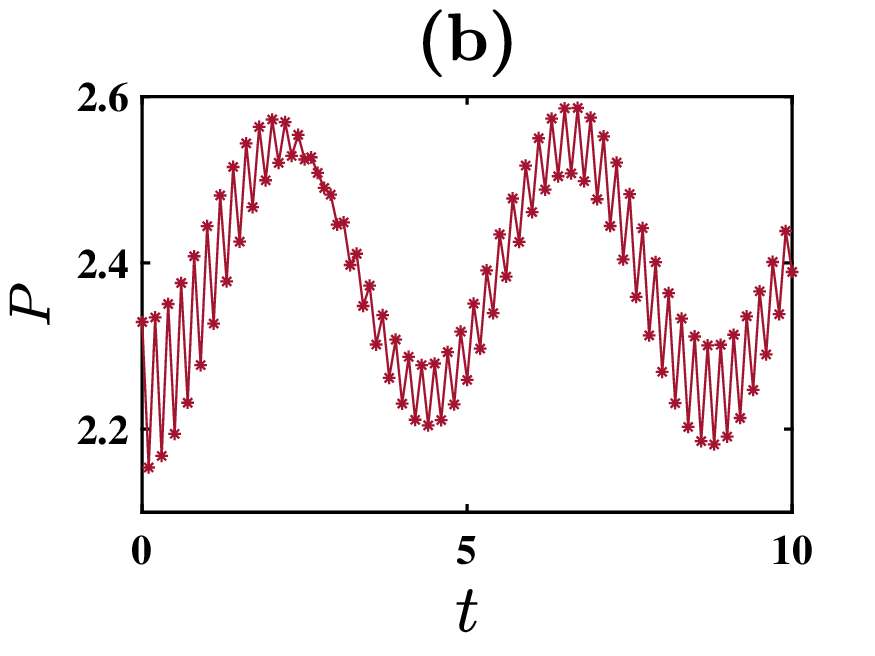}
		\includegraphics[width=0.4\textwidth]{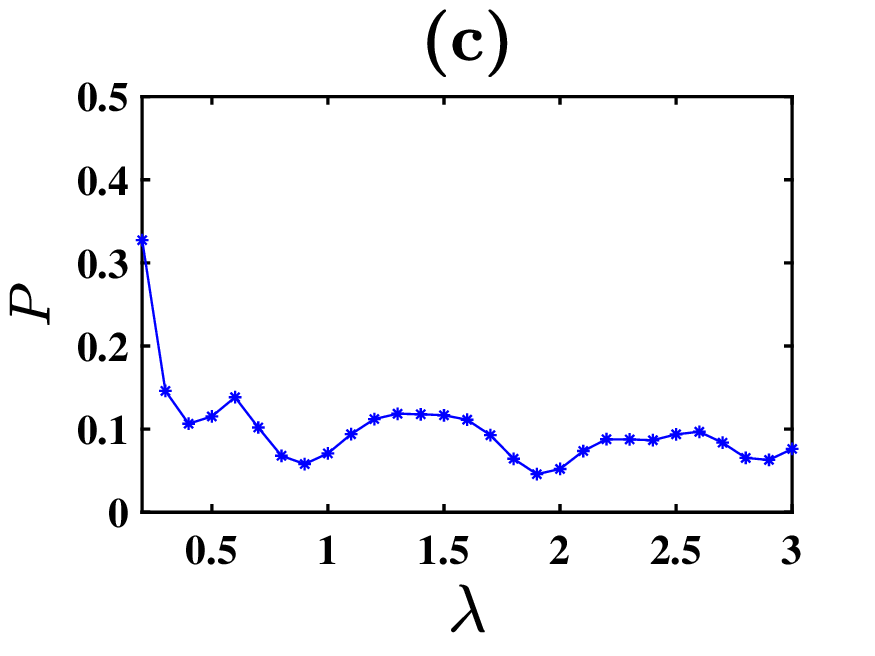}
		\includegraphics[width=0.4\textwidth]{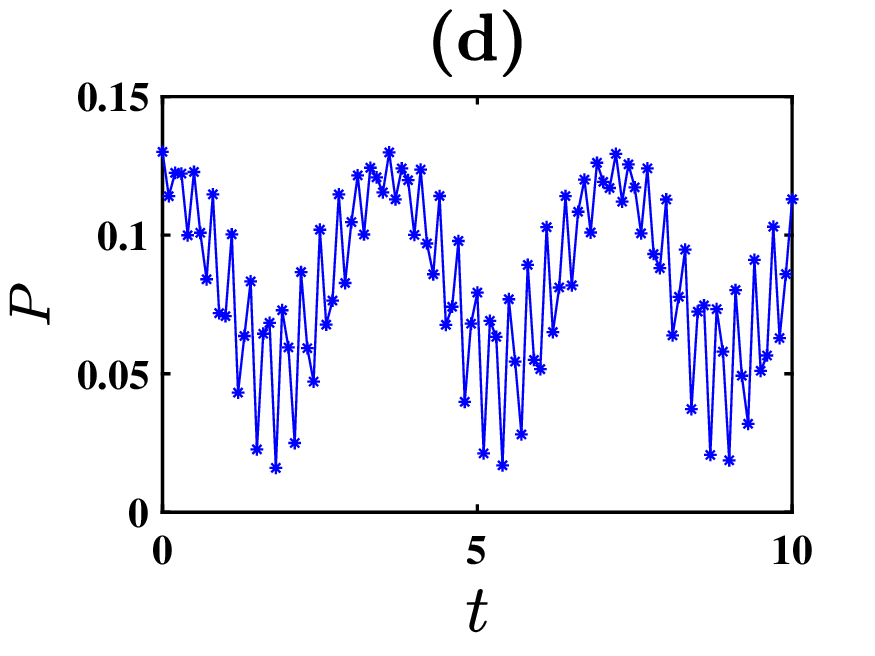}
		\caption{Variation of photon number distribution $P(n_1,n_2)$ for fixed values $\chi=0,\chi_1=\chi_2=1,n_1=n_2=10,\alpha_1=\alpha_2=1,\phi=0,\beta_1=\beta_2=0,\Delta=30,\gamma_1=\gamma_2=1/\sqrt 2$ and as a function of \textbf{(a)} $\lambda$ with $f(n_1)=f(n_2)=1$, $t=1$, \textbf{(b)} $t$ with $f(n_1)=f(n_2)=1$, $ \lambda=1$, \textbf{(c)} $\lambda$ with $f(n_1)=\sqrt{n_1}$, $f(n_2)=\sqrt{n_2}$, $t=1$, \textbf{(d)} $t$ with $f(n_1)=\sqrt{n_1}$, $f(n_2)=\sqrt{n_2}$, $\lambda=1$.}
		\label{pn}
	\end{figure}

	Photon number distribution $P(n)$ represents the probability distribution of detecting $n$ number of photons in a given electromagnetic field state. It characterizes the statistical distribution of photons in a quantum system, providing insights about the usage of different photon number states. Mathematically, it can be defined as the expectation value of a given state $\ket{\psi}$ in Fock state basis as:
	$$P(n)=|\langle n|\psi\rangle|^2$$
	This distribution is fundamental to understand the behaviour of light in various quantum optical systems and is often employed for analysing different phenomena such as thermal radiation, cavity interaction, and quantum optics experiments. For a two-mode quantized field, photon number distribution is the probability of finding $n_1$ photons in first mode and $n_2$ photons in second mode \cite{Swain:22} which can be calculated as
	\begin{align}\nonumber
		& P(n_1, n_2)\\
		&=|q_{n_1}|^2 |q_{n_2}|^2|A_1|^2+2|q_{n_1-1}|^2|q_{n_2-1}|^2|A_2|^2+|q_{n_1-2}|^2|q_{n_2-2}|^2|A_4|^2
		\label{pnd}
	\end{align}
	The simplified version of the photon number distribution \eqref{pnd} is plotted in Fig.~\ref{pn}. It is evident from the plot that the photon number distribution (PND) exhibits wave-like behaviour over time in both the linear ($f(n)=1$) and nonlinear  ($f(n)=\sqrt{n}$) cases, keeping other parameters fixed. In addition, while considering the variation of PND with respect to $\lambda$, it decreases as $\lambda$ increases for $f(n)=1$ but increases with increasing $\lambda$ for $f(n)=\sqrt{n}$.
	
	\section{Mandel's $Q_M$ parameter}
	Mandel's $Q_M$ parameter is a metric employed in quantum optics for assessing the statistical attributes of light, particularly its photon number distribution. It serves to quantify the magnitude of photon number fluctuations in a given optical state relative to a coherent state. This parameter is defined mathematically as \cite{mandel1979sub}
	$$Q_M = \frac{\langle {N}^2 \rangle - \langle N \rangle^2}{\langle N \rangle} - 1$$
	where $\langle N^2 \rangle$ represents the second moment of the photon number distribution and $\langle N \rangle$ is the average photon number.
	Mandel's $Q_M$ parameter helps to characterize and compare the statistical properties of different optical states by highlighting deviations from the behaviour of a coherent state. 
	
\begin{figure}[htb]
		\centering
		\includegraphics[width=0.4\textwidth]{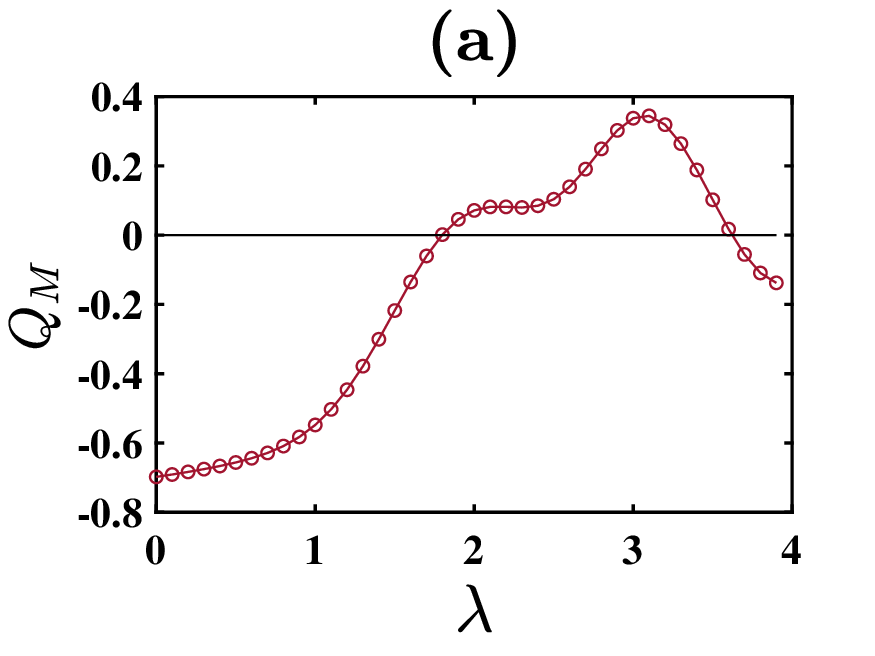}
		\includegraphics[width=0.4\textwidth]{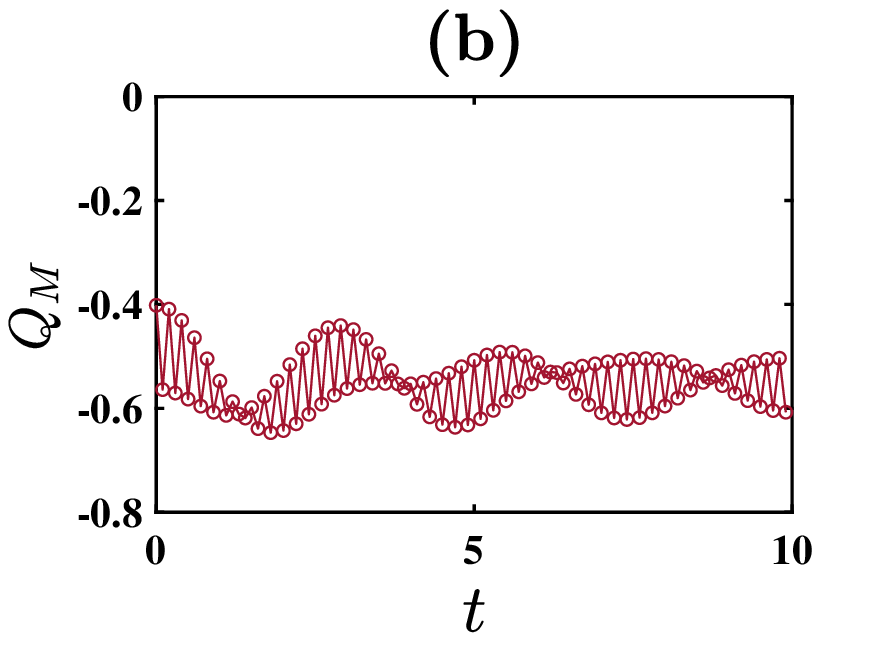}
		\includegraphics[width=0.4\textwidth]{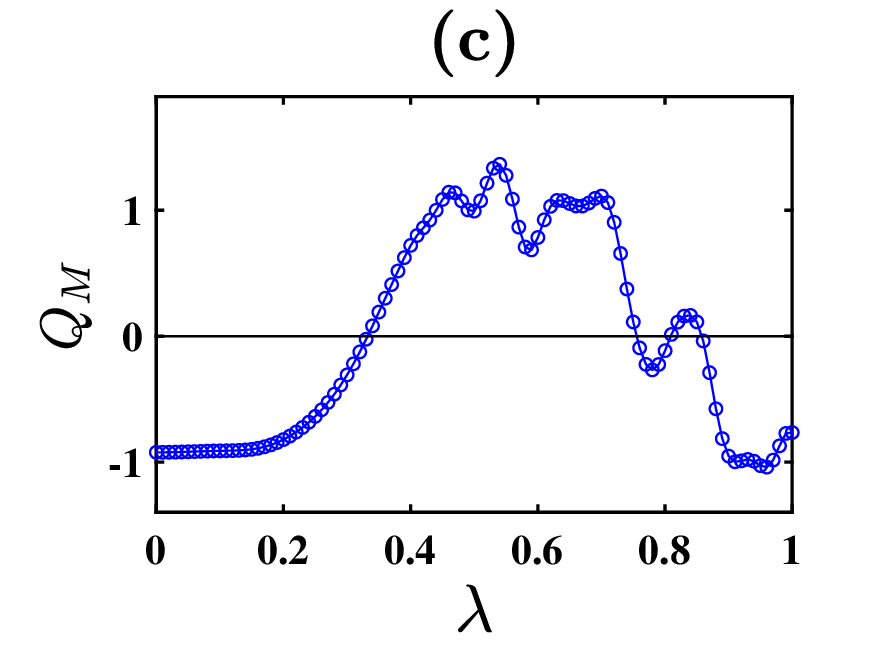}
		\includegraphics[width=0.4\textwidth]{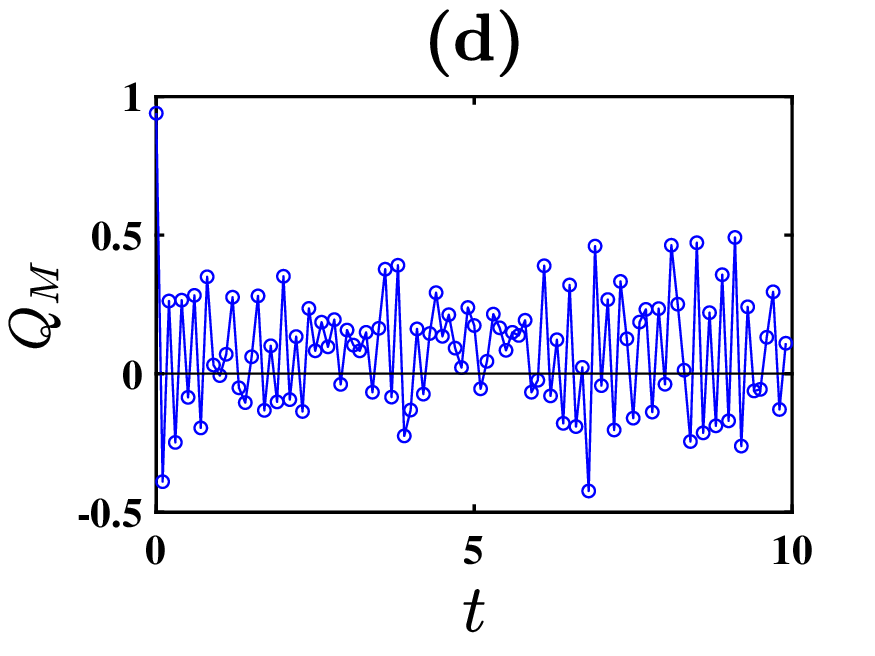}
		\caption{Variation of Mandel's $Q_M$ parameter for fixed values $\chi=0,\chi_1=\chi_2=1,\alpha_1=\alpha_2=1,\phi=0,\beta_1=\beta_2=0,\Delta=10,\gamma_1=\gamma_2=1/\sqrt 2$ and as a function of \textbf{(a)} $\lambda$ with $f(n_1)=f(n_2)=1$, $t=1$, \textbf{(b)} $t$ with $f(n_1)=f(n_2)=1$, $\lambda=1$, \textbf{(c)} $\lambda$ with $f(n_1)=\sqrt{n_1}$, $f(n_2)=\sqrt{n_2}$, $t=1$, \textbf{(d)} $t$ with $f(n_1)=\sqrt{n_1}$, $f(n_2)=\sqrt{n_2}$, $\lambda=1$.}
		\label{Q}
	\end{figure}

	A value of $ Q_M=0$ indicates that the photon number distribution is Poissonian, typifying coherent states. $Q_M>0$ suggests that the photon number statistics is super-Poissonian, indicating more photon number fluctuations than a classical-like coherent state whereas $Q_M<0$ indicates sub-Poissonian photon statistics which is often associated with squeezed states and nonclassical light. Here Mandel’s $Q_M$ function is plotted for $f(n)=1$ and $f(n)=\sqrt{n}$ and with resect to different parameters in Fig.~\ref{Q}.  
	
	We can observe that $Q_M$ becomes negative with respect to $\lambda$ and $t$ for both linear ($f(n)=1$) (in maroon) and nonlinear ($f(n)=\sqrt{n}$) (in blue) functions. Based on the nonclassicality criteria for $Q_M$, we can conclude that the quantum state under consideration exhibits nonclassical characteristics.
	\section{Second-order correlation}
	The second-order correlation function $g^2(0)$ is a key tool in quantum optics for analysing photon statistics and detecting phenomena such as photon bunching or antibunching. This second-order correlation function $g^2(0)$ for a single-mode radiation field is defined as \cite{loudon1987squeezed}\\
	\begin{equation}
		g^2(0)=\frac{\bra{\psi_\mathrm {field}}a^{\dag 2}a^2\ket{\psi_\mathrm {field}}}{\bra{\psi_\mathrm {field}}a^{\dag}a\ket{\psi_\mathrm {field}}^2}
	\end{equation}
			The value of $g^2(0)<1$ $(g^2(0)>1)$ corresponds to sub-Poissonian (super-Poissonian) statistics of the cavity mode, which is a nonclassical (classical) effect. This effect of the sub-Poissonian (super-Poissonian) statistics is often referred to as photon antibunching (bunching) \bluecite{qu2020improving}.
	
\begin{figure}[htb]
		\centering
		\includegraphics[width=0.4\textwidth]{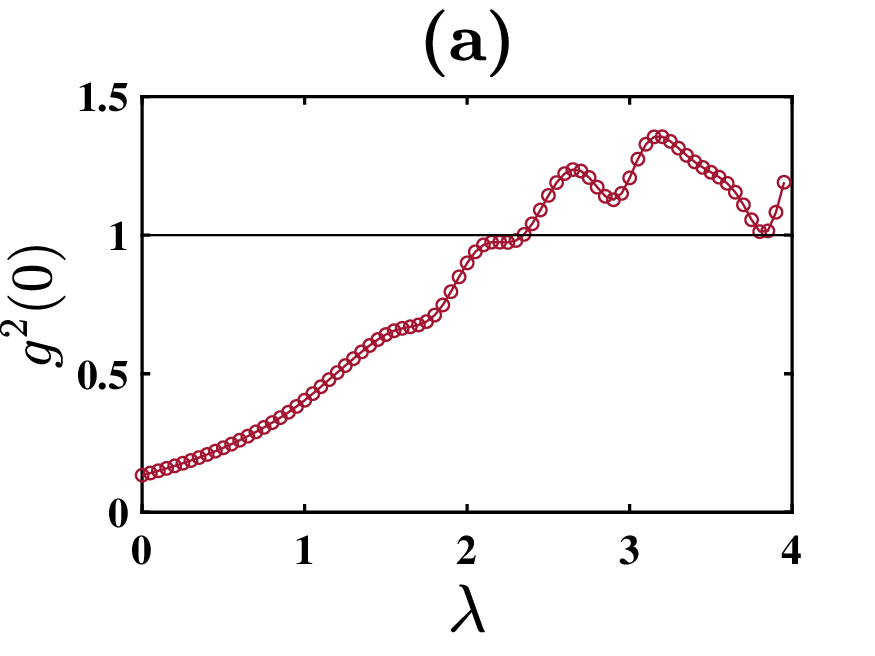}
		\includegraphics[width=0.4\textwidth]{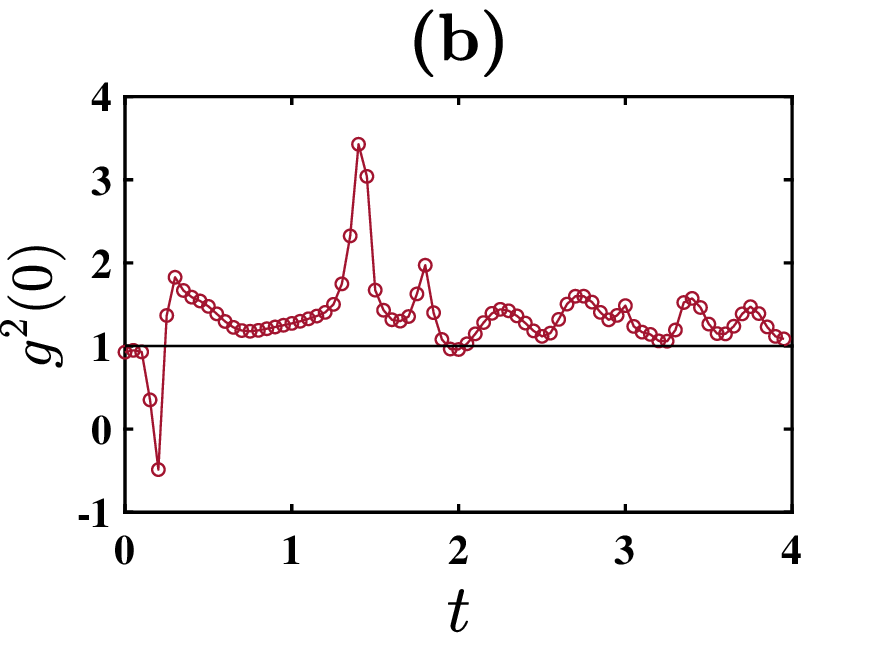}
		\includegraphics[width=0.4\textwidth]{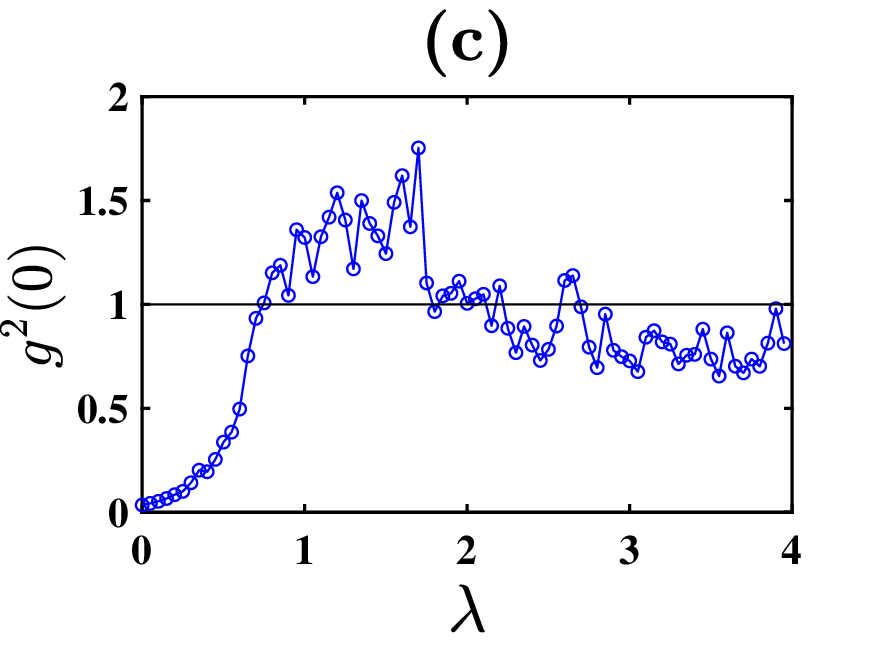}
		\includegraphics[width=0.4\textwidth]{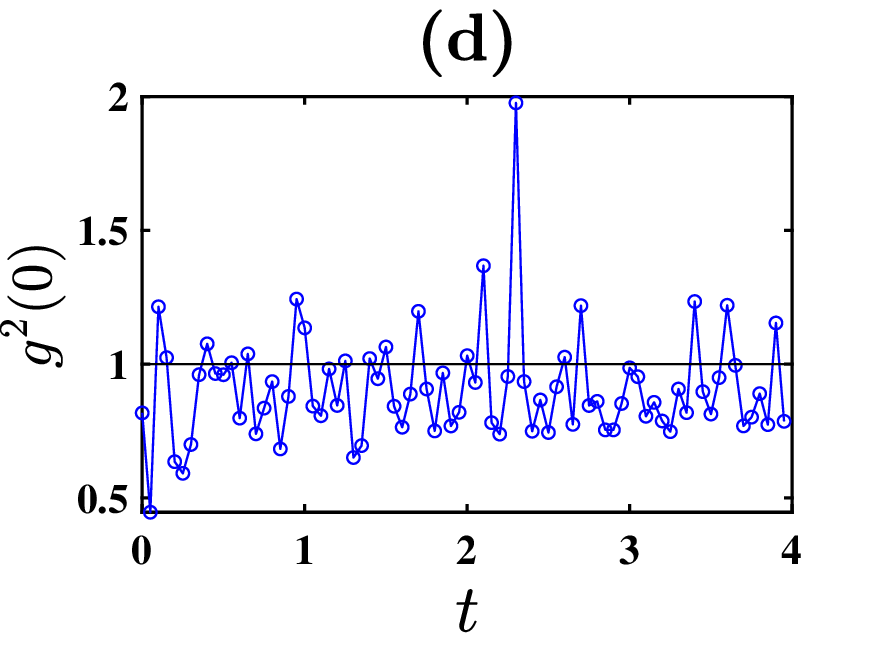}
		\caption{Variation of second-order correlation function $g^2(0)$  for fixed values $\chi=0=\chi_1=\chi_2=1$, $\alpha_1=\alpha_2=1$, $\phi=0$, $\beta_1=\beta_2=0$, $\Delta=10$, $\gamma_1=\gamma_2=1/\sqrt 2$ and as a function of \textbf{(a)} $\lambda$ with $f(n_1)=f(n_2)=1$, $t=1$, \textbf{(b)} $t$ with $f(n_1)=f(n_2)=1$,  $\lambda=1$, \textbf{(c)} $\lambda$ with $f(n_1)=\sqrt{n_1}$, $f(n_2)=\sqrt{n_2}$, $t=1$, \textbf{(d)} $t$ with $f(n_1)=\sqrt{n_1}$, $f(n_2)=\sqrt{n_2}$, $\lambda=1$.}
		\label{g2}
	\end{figure}

	We have plotted the second-order correlation function $g^2(0)$ as a function of $\lambda$ and $t$ with $f(n)=1$ in maroon and $f(n)=\sqrt{n}$ in blue color. It is seen from Fig.~\ref{g2} that when $\lambda$ and $t$ are varied, the value of $g^2(0)$ drops below one for certain combinations of the parameters. This indicates that the cavity field state is antibunched for both the linear as well as nonlinear cases. Here the correlation function $g^2(0)$ corresponding to $f(n)=\sqrt n$ shows more fluctuations.
	\section{Squeezing properties}
	In quantum mechanics, squeezing describes a phenomenon in which the uncertainty (or quantum noise) in one observable is reduced below the threshold limit at the expense of increased uncertainty in its complementary observable. This is a direct consequence of the Heisenberg uncertainty principle \cite{heisenberg1927anschaulichen}, which states that the product of the uncertainties in certain pairs of physical properties, like position and momentum, has a lower bound. In the context of quantum optics, squeezing can be applied to the field's quadrature where the uncertainty in one quadrature can be squeezed below the standard quantum noise limit while the uncertainty in the orthogonal quadrature increases \cite{walls2008quantum}.
	
	To analyze the quantum fluctuations of the field's quadrature, we consider two Hermitian operators formulated by the combinations of photon creation and annihilation operators as
	
	$$\hat{x}=\frac{{a}+{a}^{\dagger}}{2},\,\,\,\,\,\,\hat{p}=\frac{{a}-{a}^{\dagger}}{2i}$$\\
	with the commutation relation $[\hat{x},\hat{p}] = i/2$. They satisfy the Heisenberg uncertainty principle, $\langle(\Delta\hat{x})^2\rangle\langle(\Delta\hat{p})^2\rangle\geq1/16$, indicating quadrature squeezing when $\langle(\Delta\hat{x})^2\rangle<\frac{1}{4}$ or $\langle(\Delta\hat{p})^2\rangle<\frac{1}{4}$.
	It is convenient to introduce the squeezing parameters as \cite{Kumar2022}
			\begin{align*}
			s_x&=4(\Delta x)^2 – 1\\
			&= 2\langle{a}^\dag{a}\rangle+\langle{a}^2\rangle+\langle{a}^{\dag2}\rangle-
			\langle{a}\rangle^2-\langle{a}^\dag\rangle^2-2\langle{a}\rangle\langle{a}^\dag\rangle,
		\end{align*}
		\begin{align*}
			s_p&=4(\Delta p)^2 – 1\\
			&= 2\langle{a}^\dag{a}\rangle-\langle{a}^2\rangle-\langle{a}^{\dag2}\rangle+
			\langle{a}\rangle^2+\langle{a}^\dag\rangle^2-2\langle{a}\rangle\langle{a}^\dag\rangle,
		\end{align*}
		where 
		\begin{align*}
			(\Delta x)^2&=\langle x^2 \rangle-\langle x \rangle ^2\\
			(\Delta p)^2&=\langle p^2 \rangle-\langle p \rangle ^2
	\end{align*}
	\begin{figure}[htb]
		\centering
		\includegraphics[width=0.4\textwidth]{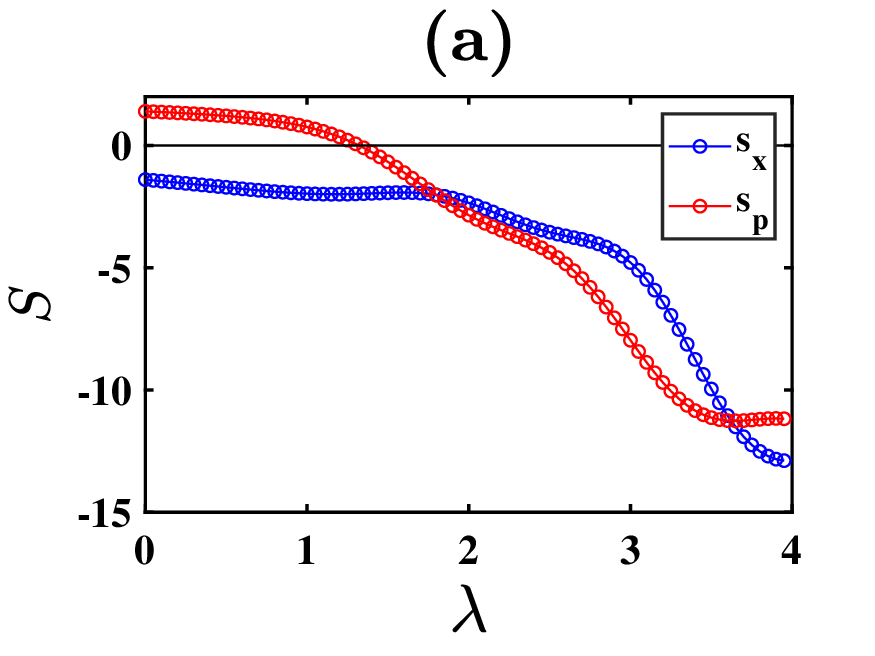}
		\includegraphics[width=0.4\textwidth]{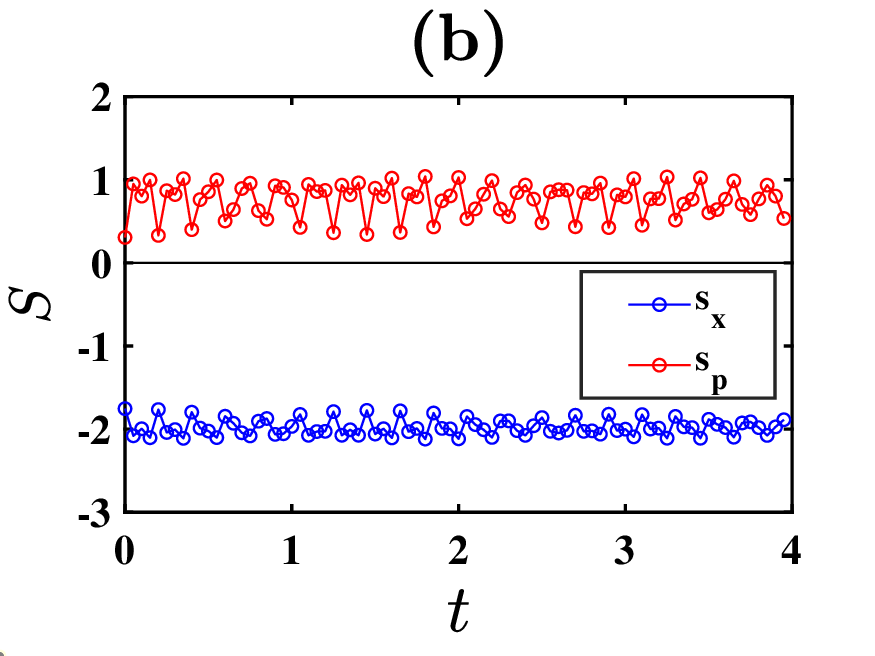}
		\includegraphics[width=0.4\textwidth]{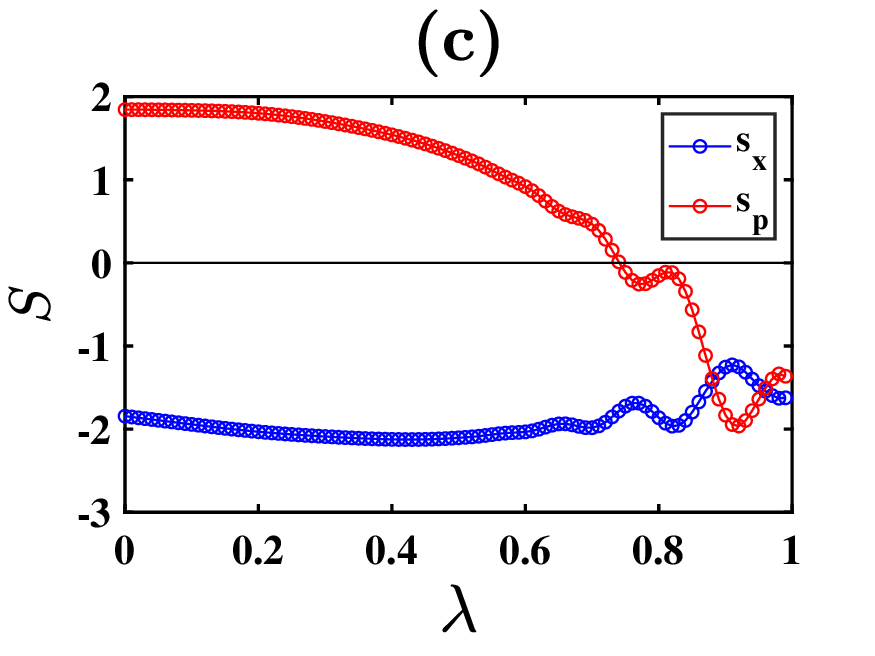}
		\includegraphics[width=0.4\textwidth]{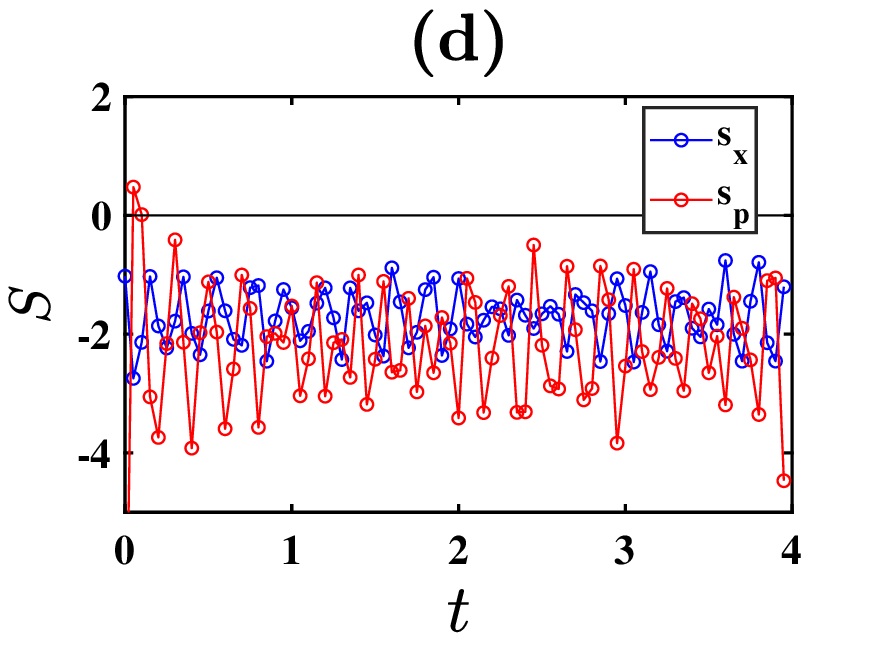}
		\caption{Variation of Squeezing properties $(s_x,\,s_p)$  for fixed values $\chi=\chi_1=\chi_2=0$, $\alpha_1=\alpha_2=1$, $\phi=0$, $\beta_1=\beta_2=0$, $\Delta=10$,$\gamma_1=\gamma_2=1/\sqrt 2$ and as a function of \textbf{(a)} $\lambda$ with $f(n_1)=f(n_2)=1$, $t=1,$ \textbf{(b)} $t$ with $f(n_1)=f(n_2)=1$, $\lambda=1$, \textbf{(c)} $\lambda$ with $f(n_1)=\sqrt{n_1}, f(n_2)=\sqrt{n_2}$, $t=1$, \textbf{(d)} $t$ with $f(n_1)=\sqrt{n_1}$, $f(n_2)=\sqrt{n_2}$, $\lambda=1$.}
		\label{S}
	\end{figure}
	Squeezed states of light are useful in quantum optics and quantum information processing as they have significant potential to improve the precision of measurements beyond the standard quantum limit, to enhance the capabilities of quantum
	devices. They are crucial in the generation of entangled photon pairs, which are essential for various quantum technologies including quantum teleportation and quantum cryptography. 
	
	We have plotted the squeezing values $S=(s_x,\,s_p)$ as a function of $\lambda$ and $t$ in Fig. \ref{S}. Here $S=(s_x,\,s_p)$ becomes negative as a function of $\lambda$ as well as $t$, indicating the nonclassical nature of the cavity field state. 
			\section{Linear Entropy}
\begin{figure}[htb]
			\centering
			\includegraphics[width=0.4\textwidth]{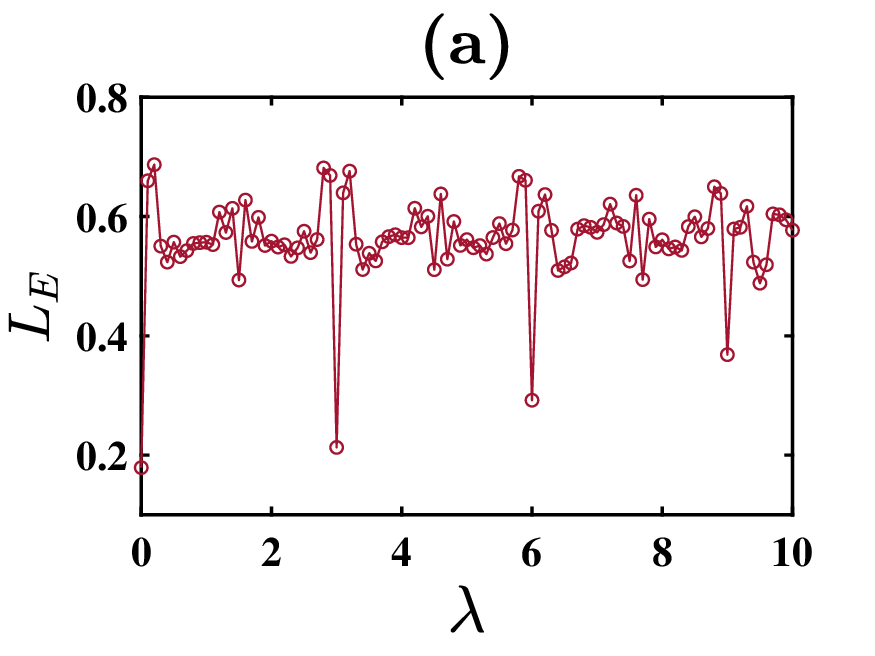}
			\includegraphics[width=0.4\textwidth]{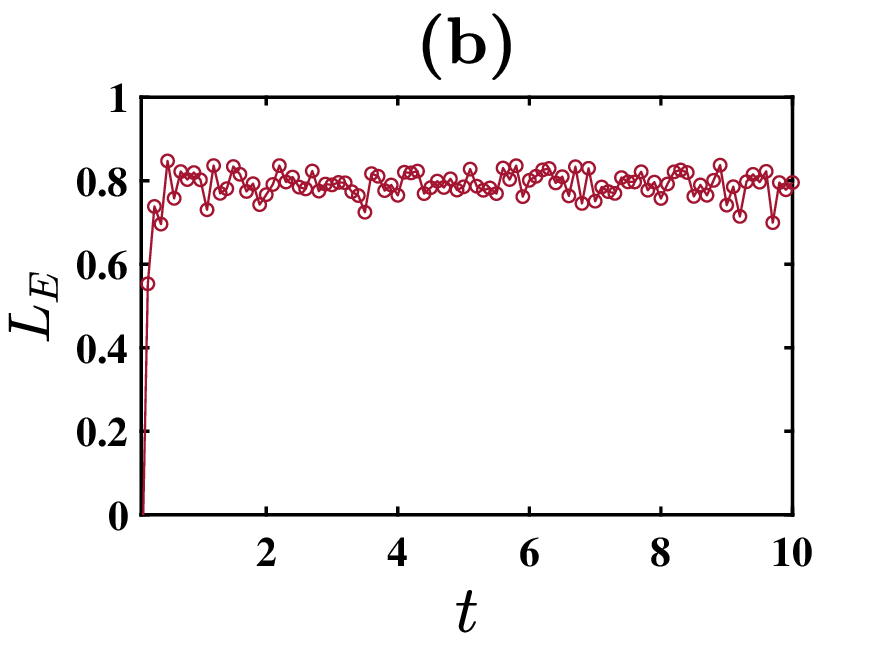}
			\includegraphics[width=0.4\textwidth]{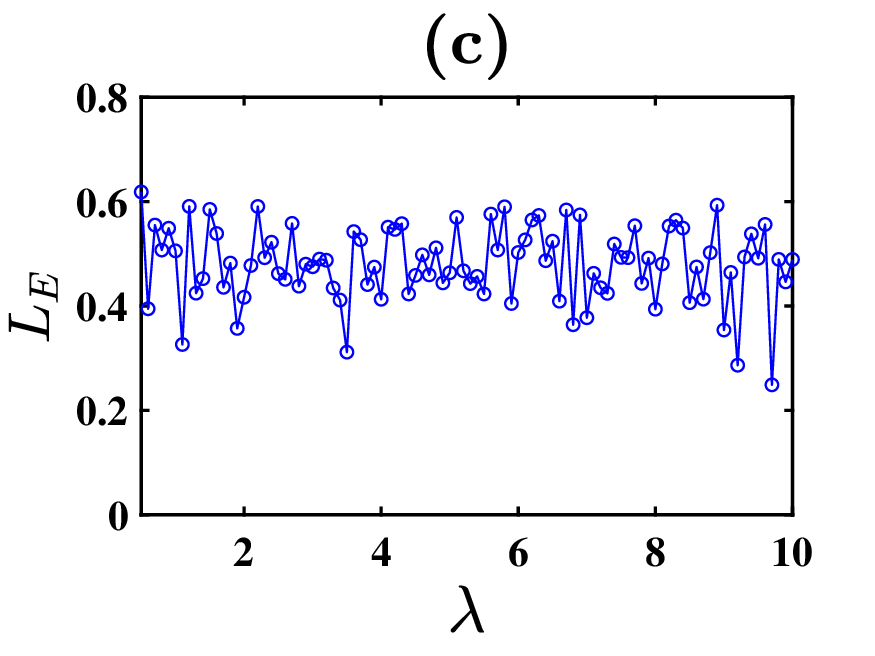}
			\includegraphics[width=0.4\textwidth]{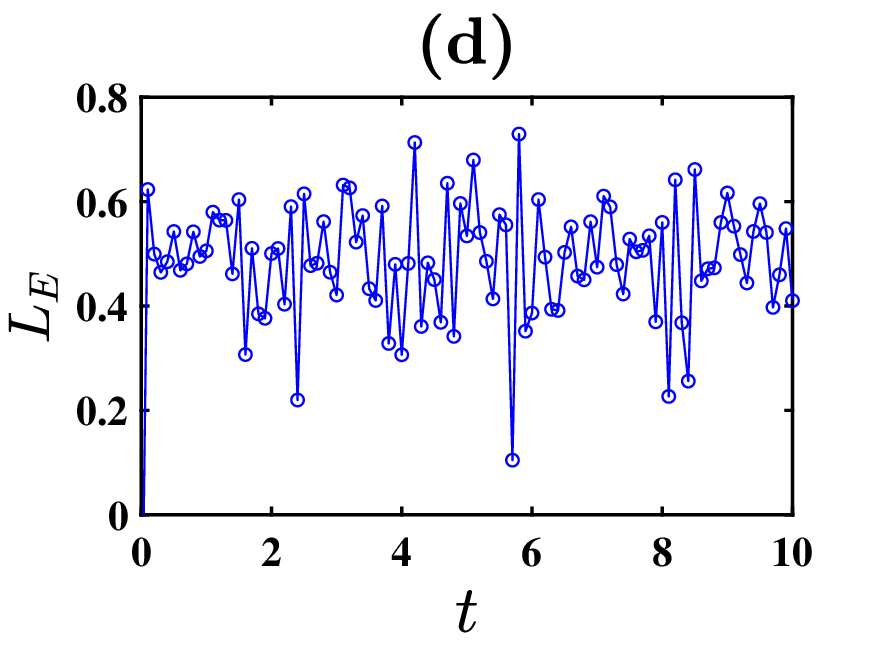}
			\caption{Variation of linear entropy $L_E(T)$  for fixed values $\chi=\chi_1=\chi_2=0$, $\alpha_1=\alpha_2=1$, $\phi=0$, $\beta_1=\beta_2=0$, $\Delta=10$,$\gamma_1=\gamma_2=1/\sqrt 2$ and as a function of \textbf{(a)} $\lambda$ with $f(n_1)=f(n_2)=1$, $t=1,$ \textbf{(b)} $t$ with $f(n_1)=f(n_2)=1$, $\lambda=1$, \textbf{(c)} $\lambda$ with $f(n_1)=\sqrt{n_1}, f(n_2)=\sqrt{n_2}$, $t=1$, \textbf{(d)} $t$ with $f(n_1)=\sqrt{n_1}$, $f(n_2)=\sqrt{n_2}$, $\lambda=1$.}
			\label{le}
		\end{figure}

		Linear entropy is a significant measure of quantum entanglement and serves as an indicator of the mixedness of a quantum state \cite{phoenix1991comment,phoenix1991establishment,ladd2002all}. For a bipartite system comprising two two-level atoms interacting with a two-mode quantized field within a Kerr-like medium, the linear entropy provides insight into the entanglement dynamics between subsystems and reveals the degree of information shared. The linear entropy $L_E(t)$ is defined as
		\begin{align*}
			L_E(t)&=1-Tr (\rho^2_{AB})
		\end{align*}
		where $\rho_{AB}$ is the reduced density matrix of the atom-atom system obtained by tracing out the field part. When $L_E=0$, the system is in a pure state $0<L_E<1$ indicates that the system is in a mixed state. At any time $t>0$, the reduced density matrix of the atom-atom system is given by 
		\begin{eqnarray}
	\rho_{AB}(t)	=\text{Tr}_\mathrm{field}[\rho(t)]=\left(
	\begin{array}{cccc}
	\rho_{11} & \rho_{12} & \rho_{13} & \rho_{14}\\
	\rho_{21} & \rho_{22} & \rho_{23} & \rho_{24}\\
	\rho_{31} & \rho_{32} & \rho_{33} & \rho_{34}\\
	\rho_{41} & \rho_{42} & \rho_{43} & \rho_{44}
	\end{array}\right),
	\end{eqnarray}	
				where 
		\begin{align*}
			\rho_{11}&=\sum_{n_1,n_2=0}^{\infty}P_{n_1,n_2}|A_1(n_1,n_2,t)|^2\\
			\rho_{22}&=\sum_{n_1,n_2=0}^{\infty}P_{n_1,n_2}|A_2(n_1+1,n_2+1,t)|^2\\
			\rho_{44}&=\sum_{n_1,n_2=0}^{\infty}P_{n_1,n_2}|A_4(n_1+2,n_2+2,t)|^2\\
			\rho_{12}&=\sum_{n_1,n_2=0}^{\infty}q_{n_1+1,n_2+1}q_{n_1,n_2}^*A_1(n_1+1,n_2+1,t)A_2^*(n_1+1,n_2+1,t)\\
			\rho_{14}&=\sum_{n_1,n_2=0}^{\infty}q_{n_1+2,n_2+2}q_{n_1,n_2}^*A_1(n_1+2,n_2+2,t)A_4^*(n_1+2,n_2+2,t)\\
			\rho_{24}&=\sum_{n_1,n_2=0}^{\infty}q_{n_1+1,n_2+1}q_{n_1,n_2}^*A_2(n_1+2,n_2+2,t)A_4^*(n_1+2,n_2+2,t)\\
			\rho_{22}&=\rho_{33}=\rho_{23},\,\,\ \rho_{12}=\rho_{13} ,\,\,\ \rho_{24}=\rho_{34} ,\,\,\ \rho_{mn}=\rho^*_{nm}
		\end{align*}
		and $P_{n_1,n_2}=P_{n_1}P_{n_2}$, $P_{n_j}=|q_{n_j}|^2$ is the probability distribution function for the field mode $j$. The linear entropy is therefore calculated as		
		\begin{align*}
			L_E&=1-\Big[\rho^2_{11}+4\rho^2_{22}+\rho^2_{44}+4|\rho_{12}|^2+2|\rho_{14}|^2+4|\rho_{24}|^2\Big]
		\end{align*}
		We have plotted the linear entropy $L_E(t)$ as a function of coupling parameter $\lambda$ and time $t$. Fig.~\ref{le} illustrates the dynamics of linear entropy as a measure of atomic entanglement for two different coupling scenarios, $f(n)=1$ in maroon and $f(n)=\sqrt{n}$ in blue color. The linear entropy, ranging from 0 to 1, quantifies the mixedness of the reduced atomic state while 0 corresponds to a pure state and 1 indicates maximum mixedness. While plotted with respect to $\lambda$, linear entropy mainly varies between $0.2$ to $0.7$ for $f(n)=1$ and $0.3$ to $0.6$ for $f(n)=\sqrt{n}$. As a function of $t$, it ranges between $0.7$ to $0.85$ ($0.2$ to $0.7$) for $f(n)=1$ ($f(n)=\sqrt{n}$). This figure reveals how the linear entropy evolves over time, highlighting the differences in atomic entanglement dynamics.
	
	\section{Conclusion}
	In this paper, we have explored the generation of a particular quantum state through the interaction of two two-level atoms and a two-mode field within an optical cavity which is embedded by a medium exhibiting Kerr nonlinearity. Initially, we have assumed that the atom is in a superposition of four primary states while the cavity itself is in a vacuum state. Following the interaction between the atoms and the cavity, we have derived the state of the cavity field by eliminating the atom's contribution from the overall state vector. We have then examined various statistical properties such as photon number distribution, Mandel’s $Q_M$ parameter, squeezing properties $s_x$ and $s_p$, and second-order correlation function $g^2(0)$. The linear entropy of two two-level atomic subsystem is calculated for the linear function $f(n)=1$ as well as the nonlinearity function $f(n)=\sqrt{n}$.
	
	Our findings revealed that Mandel’s $Q_M$ and the squeezing parameters $s_x$ and $s_p$ are displaying negative values in a particular parametric regime which indicates the nonclassical nature of the cavity field state. Moreover the second-order correlation $g^2(0)$, when examined in terms of $\lambda$ and $t$, is consistently found to be below 1, providing additional evidence of the nonclassical behavior exhibited by the cavity field. It is observed that the atomic subsystem is more entangled when $f(n)=\sqrt{n}$.
	
	The antibunched nature of the cavity field state, as observed during the investigation, signifies its potential application in single-photon sources, which are crucial in quantum communication and quantum computing protocols. These states can facilitate the generation of indistinguishable single photons, essential for tasks like quantum key distribution and linear optical quantum computing. The study of nonclassical states generated through atom-cavity interactions holds promise for a wide range of applications spanning quantum metrology, quantum communication, and quantum computing. Continued research in this area is essential for harnessing the full potential of these states and advancing quantum technologies.
	\begin{center}
		\textbf{ACKNOWLEDGEMENT}
	\end{center}
	Naveen Kumar acknowledges the financial support from the Council of Scientific and Industrial Research, Govt. of India (Grant no. 09/1256(0004)/2019-EMR-I).
	
	\begin{center}
		\textbf{Data Availability Statement}
	\end{center}
	Data generated or
	analyzed during this study are provided in full within the article.
	

\begin{thebibliography}{10}

\bibitem{Jaynes1963}
E.~T. Jaynes and F.~W. Cummings.
\newblock Comparison of quantum and semiclassical radiation theories with
  application to the beam maser.
\newblock {\em Proceedings of the IEEE}, 51(1):89--109, 1963.

\bibitem{shore1993jaynes}
B.~W. Shore and P.~L. Knight.
\newblock The {J}aynes-{C}ummings model.
\newblock {\em Journal of Modern Optics}, 40(7):1195--1238, 1993.

\bibitem{jex1992emission}
I.~Jex.
\newblock Emission spectra of a two-level atom under the presence of another
  two-level atom.
\newblock {\em Journal of Modern Optics}, 39(4):835--844, 1992.

\bibitem{bougouffa2010entanglement}
S.~Bougouffa.
\newblock Entanglement dynamics of two-bipartite system under the influence of
  dissipative environments.
\newblock {\em Optics Communications}, 283(14):2989--2996, 2010.

\bibitem{ashraf1999effects}
M.~M. Ashraf.
\newblock Effects of a phase shift on two-photon process.
\newblock {\em Optics Communications}, 166(1-6):49--55, 1999.

\bibitem{hekmatara2014sub}
H.~Hekmatara and M.~K. Tavassoly.
\newblock Sub-{P}oissonian statistics, population inversion and entropy
  squeezing of two two-level atoms interacting with a single-mode binomial
  field: intensity-dependent coupling regime.
\newblock {\em Optics Communications}, 319:121--127, 2014.

\bibitem{zidan2002influence}
A.~N. Zidan, M.~Abdel-Aty, and A.~F. Obada.
\newblock Influence of intrinsic decoherence on entanglement degree in the
  atom--field coupling system.
\newblock {\em Chaos, Solitons \& Fractals}, 13(7):1421--1428, 2002.

\bibitem{el2003entropy}
T.~M. El-Shahat, S.~A. Khalek, M.~Abdel-Aty, and A.~S.~F. Obada.
\newblock Entropy squeezing of a degenerate two-photon process with a nonlinear
  medium.
\newblock {\em Journal of Modern Optics}, 50(13):2013--2030, 2003.

\bibitem{baghshahi2015entanglement}
H.~R. Baghshahi, M.~K. Tavassoly, and M.~J. Faghihi.
\newblock Entanglement criteria of two two-level atoms interacting with two
  coupled modes.
\newblock {\em International Journal of Theoretical Physics}, 54:2839--2854,
  2015.

\bibitem{ateto2009control}
M.~S. Ateto.
\newblock Control of a nonlocal entanglement in the micromaser via two quanta
  non-linear processes induced by dynamic {S}tark shift.
\newblock {\em International Journal of Theoretical Physics}, 48:545--567,
  2009.

\bibitem{hu2008effect}
Y.~H. Hu, M.~F. Fang, J.~W. Cai, K.~Zeng, and C.~L. Jiang.
\newblock Effect of the {S}tark shift on entanglement in a double two-photon
  {JC} model.
\newblock {\em Journal of Modern Optics}, 55(21):3551--3562, 2008.

\bibitem{baghshahi2014entanglement}
H.~R. Baghshahi, M.~K. Tavassoly, and M.~J. Faghihi.
\newblock Entanglement analysis of a two-atom nonlinear {J}aynes-{C}ummings
  model with nondegenerate two-photon transition, {K}err nonlinearity, and
  two-mode {S}tark shift.
\newblock {\em Laser Physics}, 24(12):125203, 2014.

\bibitem{aspect1982experimental}
A.~Aspect, J.~Dalibard, and G.~Roger.
\newblock Experimental test of {B}ell's inequalities using time-varying
  analyzers.
\newblock {\em Physical Review Letters}, 49(25):1804, 1982.

\bibitem{nielsen2010quantum}
M.~L. Nielsen and I.~L. Chuang.
\newblock {\em Quantum Computation and Quantum Information}.
\newblock Cambridge University Press, 2010.

\bibitem{haroche2006exploring}
S.~Haroche and J.~M. Raimond.
\newblock {\em Exploring the quantum: atoms, cavities, and photons}.
\newblock Oxford University Press, 2006.

\bibitem{imamoglu1997strongly}
A.~Imamoḡlu, H.~Schmidt, G.~Woods, and M.~Deutsch.
\newblock Strongly interacting photons in a nonlinear cavity.
\newblock {\em Physical Review Letters}, 79(8):1467, 1997.

\bibitem{molmer1999multiparticle}
K.~M{\o}lmer and A.~S{\o}rensen.
\newblock Multiparticle entanglement of hot trapped ions.
\newblock {\em Physical Review Letters}, 82(9):1835, 1999.

\bibitem{poyatos1997complete}
J.~F. Poyatos, J.~I. Cirac, and P.~Zoller.
\newblock Complete characterization of a quantum process: the two-bit quantum
  gate.
\newblock {\em Physical Review Letters}, 78(2):390, 1997.

\bibitem{korashy2020dynamics}
S.~T. Korashy, T.~M. El-Shahat, N.~Habiballah, H.~El-Sheikh, and M.~Abdel-Aty.
\newblock Dynamics of a nonlinear time-dependent two two-level atoms in a
  two-mode cavity.
\newblock {\em International Journal of Quantum Information}, 18(03):2050003,
  2020.

\bibitem{abdel1992degenerate}
A.~M. Abdel-Hafez.
\newblock Degenerate and nondegenerate two-mode normal squeezing in a two-level
  atom and two-mode system.
\newblock {\em Physical Review A}, 45(9):6610, 1992.

\bibitem{miranowicz2004dissipation}
A.~Miranowicz and W.~Leo{\'n}ski.
\newblock Dissipation in systems of linear and nonlinear quantum scissors.
\newblock {\em Journal of Optics B: Quantum and Semiclassical Optics},
  6(3):S43, 2004.

\bibitem{scully1997quantum}
M.~O. Scully and M.~S. Zubairy.
\newblock {\em Quantum Optics}.
\newblock Cambridge University Press, 1997.

\bibitem{Swain:22}
S.~N. Swain, Y.~Jha, and P.~K. Panigrahi.
\newblock Two-mode photon added {S}chr\"{o}dinger {C}at states: nonclassicality
  and entanglement.
\newblock {\em Journal of the Optical Society of America B}, 39(11):2984--2991,
  Nov 2022.

\bibitem{mandel1979sub}
L.~Mandel.
\newblock Sub-{P}oissonian photon statistics in resonance fluorescence.
\newblock {\em Optics Letters}, 4(7):205--207, 1979.

\bibitem{loudon1987squeezed}
R.~Loudon and P.~L. Knight.
\newblock Squeezed light.
\newblock {\em Journal of Modern Optics}, 34(6-7):709--759, 1987.

\bibitem{qu2020improving}
Y.~Qu, S.~Shen, J.~Li, and Y.~Wu.
\newblock Improving photon antibunching with two dipole-coupled atoms in
  whispering-gallery-mode microresonators.
\newblock {\em Physical Review A}, 101(2):023810, 2020.

\bibitem{heisenberg1927anschaulichen}
W.~Heisenberg.
\newblock {\"U}ber den anschaulichen inhalt der quantentheoretischen kinematik
  und mechanik.
\newblock {\em Zeitschrift f{\"u}r Physik}, 43(3):172--198, 1927.

\bibitem{walls2008quantum}
D.~F. Walls and G.~J. Milburn.
\newblock {\em Quantum Optics}.
\newblock Springer, 2008.

\bibitem{Kumar2022}
N.~Kumar, Deepak, and A.~Chatterjee.
\newblock Nonclassical properties of a deformed atom-cavity field state.
\newblock {\em Journal of Modern Optics}, 69(18):1052--1059, 2022.

\bibitem{phoenix1991comment}
S.~J.~D. Phoenix and P.~L. Knight.
\newblock Comment on ‘‘{C}ollapse and revival of the state vector in the
  {J}aynes-{C}ummings model: an example of state preparation by a quantum
  apparatus’’.
\newblock {\em Physical review letters}, 66(21):2833, 1991.

\bibitem{phoenix1991establishment}
S.~J.~D. Phoenix and P.~L. Knight.
\newblock Establishment of an entangled atom-field state in the
  {J}aynes-{C}ummings model.
\newblock {\em Physical Review A}, 44(9):6023, 1991.

\bibitem{ladd2002all}
T.~D. Ladd, J.~R. Goldman, F.~Yamaguchi, Y.~Yamamoto, E.~Abe, and K.~M. Itoh.
\newblock All-silicon quantum computer.
\newblock {\em Physical Review Letters}, 89(1):017901, 2002.

\end{thebibliography}

\end{document}